\documentclass[conference]{IEEEtran}
\IEEEoverridecommandlockouts
\usepackage{cite}
\usepackage{amsmath,amssymb,amsfonts}
\usepackage{algorithmic}
\usepackage{graphicx}
\usepackage{textcomp}
\usepackage{xcolor}
\usepackage{hyperref}
\usepackage{orcidlink}

\def\BibTeX{{\rm B\kern-.05em{\sc i\kern-.025em b}\kern-.08em
    T\kern-.1667em\lower.7ex\hbox{E}\kern-.125emX}}

\usepackage{float}

\usepackage{enumitem}
\newlist{questions}{enumerate}{2}
\setlist[questions,1]{leftmargin=*, label=\textbf{RQ\arabic*.},ref=RQ\arabic*}

\usepackage{tabularx}

\newcommand{\linebreakand}{%
  \end{@IEEEauthorhalign}
  \hfill\mbox{}\par
  \mbox{}\hfill\begin{@IEEEauthorhalign}
}

\begin{document}

\title{Mind the Ethics! The Overlooked Ethical Dimensions of GenAI in Software Modeling Education}

\hypersetup{pdfborder={0 0 0}}

\author{\IEEEauthorblockN{Shalini Chakraborty\orcidlink{0000-0002-9466-3766}}
\IEEEauthorblockA{\textit{Software Engineering Group} \\
\textit{University of Bayreuth}\\
Bayreuth, Germany \\
s.chakraborty@uni-bayreuth.de}

\and
\IEEEauthorblockN{Lola Burgue\~{n}o\orcidlink{XXXX-XXXX-XXXX-XXXX}}
\IEEEauthorblockA{\textit{ITIS Software} \\
\textit{University of Malaga}\\
Malaga, Spain \\
lolaburgueno@uma.es}
\and
\IEEEauthorblockN{Nathalie Moreno\orcidlink{XXXX-XXXX-XXXX-XXXX}}
\IEEEauthorblockA{\textit{ITIS Software} \\
\textit{University of Malaga}\\
Malaga, Spain \\
nmv@uma.es}
\and
\IEEEauthorblockN{Javier Troya\orcidlink{XXXX-XXXX-XXXX-XXXX}}
\IEEEauthorblockA{\textit{ITIS Software} \\
\textit{University of Malaga}\\
Malaga, Spain \\
jtroya@uma.es}
\and
\IEEEauthorblockN{Paula Mu\~{n}oz\orcidlink{XXXX-XXXX-XXXX-XXXX}}
\IEEEauthorblockA{\textit{Independent} \\
Malaga, Spain \\
}
}

\maketitle
\begin{abstract}
Generative Artificial Intelligence (GenAI) is rapidly gaining momentum in software modeling education, embraced by both students and educators. As GenAI assists with interpreting requirements, formalizing models, and translating students' mental models into structured notations, it increasingly shapes core learning outcomes such as domain comprehension, diagrammatic thinking, and modeling fluency without clear ethical oversight or pedagogical guidelines. Yet, the ethical implications of this integration remain underexplored.
In this paper, we conduct a systematic literature review across six major digital libraries in computer science (ACM Digital Library, IEEE Xplore, Scopus, ScienceDirect, SpringerLink, and Web of Science). Our aim is to identify studies discussing the ethical aspects of GenAI in software modeling education, including responsibility, fairness, transparency, diversity, and inclusion among others.
Out of 1,386 unique papers initially retrieved, only three explicitly addressed ethical considerations. This scarcity highlights the critical absence of ethical discourse surrounding GenAI in modeling education and raises urgent questions about the responsible integration of AI in modeling curricula, as well as it evinces the pressing need for structured ethical frameworks in this emerging educational landscape. 
We examine these three studies and explore the emerging research opportunities as well as the challenges that have arisen in this field.

\end{abstract}


\section{Introduction}
\label{sec:intro}
Generative Artificial Intelligence (GenAI) has rapidly established itself as a transformative tool across multiple fields, particularly within education. Since the launch of OpenAI’s ChatGPT \cite{roumeliotis2023chatgpt}, this technology has gained popularity among students, providing assistance in various daily tasks ranging from grammar correction and sentence formation to coding and solving of complex assignments. Its role in both educational and non-educational settings has shown benefits (such as enhanced accessibility and immediate feedback \cite{adeshola2023opportunities}) as well as concerns (e.g., potentially increased dependency on AI tools \cite{castillo2023effect, li2024mitigating,camara2023assessment}). 

In the field of software engineering, numerous studies have examined the impact of AI on the discipline's education. In some of them, concerns both from educators as well as students have been shared such as reducing the emphasis on manual coding~\cite{daun2023chatgpt}, the subjective assessment of the correctness of AI-generated responses~\cite{jalil2023chatgpt}, and the fact that excessive reliance on AI could hinder their learning~\cite{silva2024chatgpt}.

Software modeling represents an interpretive and creative practice~\cite{Chaaben25} within software engineering. Among the various modeling approaches, the Unified Modeling Language (UML) remains a widely adopted standard in both educational and professional contexts. As in other domains, GenAI is beginning to reshape how UML-based modeling is taught and applied in practice.
While much attention has been given to the capabilities and limitations of AI in general, its long-term ethical implications in software modeling education, particularly those introduced by GenAI, remain largely unexamined. With GenAI tools now assisting in tasks such as translating requirements into models and automating diagram construction~\cite{burgueno21NLP,Rocco21,WeyssowSS22,camara2023assessment,roumeliotis2023chatgpt,ChaabenBS23}, there is a growing need to critically evaluate their impact on student learning, autonomy, and modeling integrity. Current works usually leave the ethical issues of GenAI out of their discussions.

%


In this paper, we investigate the ethical dimensions associated with the use of GenAI in software modeling education. We conduct a systematic review of the literature to thoroughly identify the state of the art and examine the extent to which ethical considerations are being addressed or overlooked in current research. To our surprise, we only found three papers (aka primary studies) that explicitly discuss ethics in their educational modeling context. We have analyzed these three studies and discuss the opportunities for new lines of research and the challenges identified in the area.

The remaining of the paper is structured as follows.
In Section~\ref{sec:method} we outline the search methodology. Section~\ref{sec:PrimaryStudies} describes the three primary studies identified by our search. Section~\ref{sec:discussion} discusses these papers based on our analysis of the ethical dimension. In Section~\ref{sec:rw} we present the related work and finally, we conclude the paper in Section~\ref{sec:conclusions}.

\begin{table*}[t]
\centering
\begin{tabular}{lccccccc}
\hline
\textbf{Search Engine} & ACM & Elsevier (Science Direct) & IEEE Explore & Scopus & SpringerLink & Web of Science & \textbf{Total} \\
\textbf{Num papers}    & 925 &  4,069 &  39    & 710   &  156            &  8 & \textbf{5,907} \\\hline
\end{tabular}
\caption{Papers retrieved by search engines}
\label{tab:paperPerEngine}
\end{table*}

\section{Review Method}
\label{sec:method}

To collect papers related to the ethical aspects of the use of GenAI in software modeling education, we followed a structured method partially inspired by the guidelines of Kitchenham~\cite{Kitchenham04} and Webster et al.~\cite{Webster02}. We explain in the following the methodology that we followed.

\subsection{Inclusion and exclusion criteria}
\label{subsec:inclusionExclusionCriteria}

The inclusion criteria define the scope of our systematic literature review and enable us to identify papers of interest. We focus on those papers that:

\begin{itemize}
    \item Are written in English, peer-reviewed, published between Jan 2022 and Feb 2025, belong to the Computer Science or Engineering fields, and are research papers with 4 or more pages (i.e., no editorials, invited keynotes, etc.).
    \item Focus on the areas of:
    \begin{itemize}
        \item Empirical studies for academia regarding software modeling with student participants,
        \item Approaches and/or tools for students where ethical aspects are covered, regardless of whether the paper has or has not have evaluated the approach/tool with student participants (i.e., its design takes into consideration ethical aspects despite the lack of evaluation of these),
        \item Methodologies,
        \item Position / vision papers that address ethical aspects in modeling education.
        \end{itemize}
\end{itemize}

Knowing that we could find papers that address the topic as their core contribution and papers whose main contribution is not ethics but they address this aspect as part of it, we decided to include both kinds.

Exclusion criteria enabled us to filter out irrelevant papers. We decided to exclude papers that do not focus on software modeling, that are secondary studies (surveys, systematic literature reviews, and mapping studies), or those papers that only mention ethic-related topics only as future work. We also excluded venues of well-known questionable or predatory practices. 

\subsection{Data sources}

We searched in six databases: ACM, Elsevier (Science Direct), IEEE Explorer, Scopus, SpringerLink and Web of Science. We did not use DBLP since it only allows the search in the title.

\subsection{Search Strategy and Paper Selection}

We defined the following search string:

\noindent\fbox{\begin{minipage}{\columnwidth}
\noindent\emph{(*GPT* OR LLM OR "foundation model") AND ("software model*" OR UML OR "conceptual model*" OR metamodel OR "meta-model" OR "conceptual schema" OR "domain model" OR "structural model" OR "behavioral model" OR "system model" OR "design model" OR "architectural model") AND (Ethic* OR responsibl* OR fair* OR explainab* OR transparen* OR bias OR priva* OR safety OR inclusiv* OR diversity OR equal* OR complian* OR sustainab*) AND ("education" OR "academia" OR student*)}
\end{minipage}}
\vspace{+0.15cm}

Wildcards (*) stand for any substring. For instance, \emph{model*} could match with: model, modeling, modelling, etc.\footnote{Please note that not all search engines support this type of queries. When queries containing wildcards were not supported, we adapted the string to the particular search engine. In addition, Elsevier limits the number of logical operators that can be used in a query hence we had to run multiple queries and merge the results.}. We searched in the whole document, i.e., title, abstract, keywords and full text. When filtering by research discipline was possible, we filtered by research articles within a computer science and/or engineering collection.

Table~\ref{tab:paperPerEngine} presents the number of papers retrieved by each search engine and the total number of papers. 

After removal of duplicates, we obtained 1,387 papers.

As part of our protocol, all authors met, discussed, defined, and agreed on the fact that papers should be included and discarded according to the criteria presented in Subsection~\ref{subsec:inclusionExclusionCriteria}.

Before splitting the papers and proceeding to manually checking whether the papers met the criteria, in order to ensure that the criteria was understood equally by the authors and did not lead to subjective interpretations, three authors met online to review the first eight papers based on the inclusion-exclusion criteria. The three authors examined the title and abstract, and when uncertain, they checked the full text for relevance to our search terms. In this pilot round, we did not conduct a full reading of the papers but instead focused on identifying mentions of our search terms and assessing their relevance to our study. All three authors reached a unanimous decision on the inclusion or exclusion of these eight papers.
Then, 10\% of the remaining papers (i.e., 70 papers) were randomly selected and the three authors independently assessed whether they should be included or excluded.\footnote{In absolute values, there was disagreement in 2 papers out of the 70 papers.}

To properly measure the agreement among the three authors, we applied the Intraclass correlation coefficient (ICC). ICC was chosen because it works for 2 or more raters, and both for continuous and ordinal data. The configuration we used is: two-way mixed, single score ICC(A,~1) check for absolute agreement. The result was $0.797654$, which means that there was a \emph{good agreement}.

Finally, those papers for which there was disagreement were discussed, and the remaining papers were split into three. Each chunk was assigned to one author to perform the inclusion/exclusion.

If during the individual exclusion of papers the authors had doubts about whether certain papers should be included or excluded, they marked them as such and the papers were discussed among the authors.

After performing inclusion-exclusion on all papers we ended up with three papers.
We will refer to the papers found as \emph{primary studies}.
Given the small number of final papers, one author\footnote{One of the three authors who participated in the literature review.} conducted backward snowballing \cite{wohlin2014guidelines} on the primary studies to identify any additional relevant work. However, this process did not yield any new papers\footnote{The paper by Kasneci et al. \cite{kasneci2023chatgpt}, which is cited in two of our primary studies \cite{DeBari24, Morales2023}, was a close candidate to be added. Although it discusses AI in education, it does not focus or refer to modeling, and therefore it was not included}.
All extracted data from the literature review and snowballing process is included in the supplementary material\footnote{https://doi.org/10.5281/zenodo.16185787}.



\subsection{Data extraction}
\label{subsec:dataextraction}
The primary studies are listed in Table~\ref{tab:finalpapers} with the corresponding years and ethical implications mentioned, i.e., the specific keywords from our search string that each paper matched.

The same three authors conducted a close reading of the three selected papers and summarized their findings. 
In section~\ref{sec:PrimaryStudies} we present a short summary of each primary study focusing on their contribution on software modeling and ethical aspects. 

As a sanity check, we shared with the authors of the primary studies our summaries of their works to confirm that the information collected was correct. Some minor changes were proposed and integrated.

\begin{table}
  \caption{Primary studies}
  \label{tab:finalpapers}
  \vspace{-0.25\baselineskip}
  \setlength{\tabcolsep}{5pt}
  \centering
  \begin{tabular}
  {|l|l|l|}
    \hline 
    \textbf{Authors} & \textbf{Year} & \textbf{Ethical\newline Implications} \\
    \hline

    \textsc{De Bari et al.~\cite{DeBari24}} & 2024 & fair, bias, diverse \\
    \hline 
    
    \textsc{Morales et al.~\cite{Morales2023}} & 2023 & fair, bias, diversity, priva*, equal*\\
    \hline 

    \textsc{Bucchiarone et al.~\cite{Bucchiarone24}} & 2024 & fair, bias, inclusiv*, equal*, diverse \\
    \hline

  \end{tabular}
\end{table}



\subsection{Limitations and Threats to Validity}
\emph{External Validity. }
A threat concerning external validity is tied to the selection of primary studies. This review includes three papers; however, some relevant studies may have been overlooked, either because they were not published in academic venues or did not align with our search criteria. To address this issue, we employed an extensive set of keywords using wildcards to broaden the scope of our search. Additionally, the search was constrained to peer-reviewed sources to ensure the reliability of the included works, though this may have excluded some papers not indexed in these databases. To further reduce this risk, we supplemented our search with a manual snowballing process by checking the citations from the final set of primary studies.

\emph{Internal Validity. }The search procedure was conducted systematically using automated queries, which minimizes the risk of human error. The exclusion phase—arguably the most sensitive part of the protocol was performed manually and resulted in three final papers. To reduce potential subjectivity during this phase, the three authors first reviewed a shared subset of papers to align their understanding of the inclusion and exclusion criteria. Agreement was assessed using the Interclass Correlation Coefficient (ICC). Afterward, each author independently applied the criteria to different subsets of the remaining papers. 
A potential issue could be limiting human analysis to titles and abstracts which may lead to overlooking relevant papers, particularly in cases of multi-contribution papers where ethics may not be explicitly mentioned in the abstract. To mitigate this, our exclusion phase was conducted manually by three authors working collaboratively. Whenever uncertainties arose, we discussed our findings collectively to ensure careful decision-making. In addition, we employed the \emph{Find and Search} function within abstracts and full texts to systematically check for our predefined set of keywords from the literature review.Another potential threat lies in the interpretation of the final three primary studies, where misinterpretation or omission could lead to inaccuracies. This risk was mitigated by having all three authors carefully read and validate the findings, as detailed in subsection~\ref{subsec:dataextraction}.

\section{Primary Studies}
\label{sec:PrimaryStudies}

This section presents a summary of the three primary studies identified in our systematic literature review.
\begin{itemize}
    \item De Bari et al.~\cite{DeBari24}: In this paper, the authors assess the ability of LLM agents to generate UML class diagrams within the context of requirements modeling in software engineering. Their focus is on evaluating the usefulness of LLM-generated solutions in an educational setting, specifically, how these solutions compare to those created by human students, and whether LLMs can effectively generate example solutions for instructional purposes. The study involves an experiment where both humans and an LLM solve the 20 modeling tasks (class diagrams), and the resulting diagrams are evaluated for correctness. The findings indicate that LLM-generated models tend to have significantly more semantic errors and greater textual deviation from reference solutions, although no notable differences are found in syntactic or pragmatic quality. \\   
    While selecting the modeling problems, the authors emphasize both a fair and diverse selection. Regarding fairness, they state that ``\textit{The constraints for the UML class diagram exercise were carefully chosen to create a challenging yet fair environment that mimics real-world conditions.}'' To ensure diversity, they say that they ``\emph{collected 20 exercises from a diverse set of web sources and compared the models generated by a human and an LLM solver in terms of syntactic, semantic, pragmatic correctness, and distance from a provided reference solution}''.
    However, despite these intentions, the authors acknowledge a key limitation: \textit{``All the exercises were solved by a single author of this manuscript, introducing a bias in the correctness of the exercises."} This bias potentially undermines the intended fairness and diversity of the exercise selection.

    \item Morales et al.~\cite{Morales2023}: In their paper, the authors explore their impact on model-driven software engineering education from both educator and student perspectives, outlining key opportunities and risks. It is a short, discussion paper where the authors laid certain advantages and disadvantages of integrating GenAI in modeling education.
    \\   
    In their discussion, the authors emphasize that the integration of GenAI into education brings notable risks related to fairness and potential biases. One of the primary concerns raised is the issue of data protection: ``\textit{privacy of educator and student data is not guaranteed when using generative AI tools}''. In addition to privacy, the authors point out challenges in ensuring fairness and equal treatment for all students, noting that ``\textit{Typically, generative AI tools establish for every individual user a particular context and way to response through the interaction history. Thus, it is difficult to set up equal conditions for all students}''. These concerns collectively highlight the need for careful consideration of ethical implications when integrating GenAI into educational settings.
    
    \item Bucchiarone et al.~\cite{Bucchiarone24}: In their paper, the authors present a GenAI-driven approach, combined with open educational resources (OERs), to support educators in designing lesson plans for complex subjects such as Model-Driven Engineering (MDE). The method helps teachers define clear learning objectives and recommends and generates specific learning activities by either leveraging existing educational resources or by generating new resources. At the core of this approach is a metamodel that generalizes the concept of activity, allowing the formalization of a high level methodology to streamline the lesson plan deigning process. A preliminary evaluation was conducted with 16 educational experts (9 female, 7 male) from universities and professional training institutions across diverse disciplines.

    While the authors acknowledge potential issues with GenAI—particularly concerning unfairness, inclusivity, equality and biases, they emphasize these concerns explicitly: ``\textit{One of the main challenges is the potential for biases in the AI’s responses, which can arise from the data the model was trained on. This can lead to unfair or unbalanced outputs, which may not represent diverse perspectives}''. They further analyze the impact of AI in education about inclusivity, as the content generated might not be equally accessible or relevant to all students, or, on the other side, might broaden the accessibility of high quality education.

    In response to these concerns, the authors actively promote diversity through their proposed approach. As they note, ``\textit{The lesson selection pathway offers users the flexibility to choose between retrieving a lesson from an Open Educational Resources (OER) database, crafting a lesson manually, or employing AI tools for lesson generation. This variety caters to diverse needs and preferences, ensuring that the system is adaptable to various educational contexts}''.
\end{itemize}

\section{Results and Discussion}
\label{sec:discussion}
The results of our systematic literature review were both revealing and concerning. Out of 1,387 papers, only three met our inclusion criteria, highlighting a striking lack of attention to the ethical implications of integrating GenAI into software modeling education. Of the three papers, two \cite{Bucchiarone24,DeBari24} specifically address dimensions of fairness, bias, and diversity in the context of using AI within their tools or experiments. The third paper \cite{Morales2023} provides a broader discussion on integrating GenAI into education, highlighting some general ethical risks associated with its use.

\subsection{Ethical Risks}
Apart from our final three papers, many papers mention ethics only in passing, often limited to a single line noting ethics approval for studies involving students or educators—such acknowledgments fall short. What remains largely unexplored is the deeper question of how these AI-driven experiments influence students' learning, autonomy, and understanding. The three papers we did include offer a glimpse into a few negative ethical considerations, underscoring the urgent need for more robust discourse in this area.
For example, Bucchiarone et al.~\cite{Bucchiarone24} mention that on the negative side, participants reported that the open educational resources (OERs) may not
always be updated or maintained, leading to outdated information, \textit{``OERs are not updated regularly, which is a problem in many
fields. You have to review the content thoroughly to ensure
it is still relevant and pertinent.'' } Participants noted limited search results and repeated keywords due to database normalization.
%
%

None of the three papers offered a dedicated discussion on the topic of ethics. While some concerns related to bias, unfairness, or inclusion were mentioned, an explicit reflection on the ethical responsibilities of applying GenAI in education was notably absent. Morales et al.~\cite{Morales2023} raise an important issue regarding the nature of generative AI tools in modeling contexts: such tools typically assume the input prompt to be correct and complete, and inherently try to learn from it. As they point out, \textit{``It cannot determine whether there is contradictory or missing information, or whether there are errors that should be corrected.''} In modeling, this serves an ambiguity in the diagram if generated by AI with incomplete prompt, leading to limited information for students to adapt for later modeling. 

In \cite{daun2023chatgpt}, the authors emphasize the increasing importance of software design and architecture as AI-generated code becomes integrated into the SE curriculum. As they state, \emph{``As AI algorithms can generate code based on natural language inputs, software engineers will require a more profound knowledge of software system design and architecture. This includes understanding how to properly define the problem and requirements...''}. If, however, the models—which are the primary elements of the design and architecture phase—are themselves being generated by AI, the importance of integrating ethical aspects into the curriculum becomes even more critical. Students may use prompts to define problems and requirements, while AI systems generate models from those prompts, thereby forming the foundation of the developed software. Programming built upon such models will be more susceptible to bias than ever before.

In programming contexts, studies are already emerging that highlight concerns about overreliance on AI. For instance, in \cite{park2025evaluating}, survey results show that 40.6 percent of students expressed uncertainty about responsible AI usage and potential overdependence. The authors further report that a significant performance decline in AI-free tasks suggests a dependency on Copilot, potentially hindering the development of essential independent problem-solving skills. These risks are particularly evident in programming, where biases and errors often become visible at runtime. In contrast, with modeling, immediate runtime feedback is not always available, posing an even greater threat: biases may not only influence students’ thinking but also propagate unnoticed into the developed software, ultimately affecting end results without being detected.

\subsection{Ethical Influence in Cognition}
Another significant gap in modeling education lies in understanding the cognitive effects of using GenAI. Modeling is inherently a creative and cognitively demanding task, and students already face numerous challenges including time constraints, ambiguous expectations, and limited feedback~\cite{chakraborty2023we}. When GenAI-generated feedback or modeling instructions are introduced without proper rules and regulation, there is a greater risk that students may receive misleading or inaccurate guidance, further compounding their difficulties and potentially distorting their learning process. 
Morales et al.~\cite{Morales2023} emphasize that ``\textit{a strict and correcting attitude is often needed, and replies should be given that explicitly point to incorrect input, where correctness is determined through a previously explicitly stated context}''. For instance, if a student is expected to apply the correct syntax of a language they have been taught, GenAI tools should not adapt to the student's mistakes or reinterpret the language. Instead, ``\textit{the language has to be regarded as fixed and generative AI tools should not try to learn a variation of the language presented by student faults}''.
Moreover, the authors highlight that obtaining meaningful feedback from GenAI tools often depends on highly specific prompting: \textit{``A user needs to make sure to give the proper prompt to get a suitable response. For instance, if you want feedback, you need to explicitly ask for it.''} This reliance on precise input adds to the challenge of reproducibility, as GenAI outputs can vary depending on the interaction context and inherent randomness: \textit{``Through an established interaction context and through the degree of randomness and decision freedom allowed for generative AI tools, tool responses may sometimes be irreproducible.''} While GenAI tools may perform well in general-purpose scenarios, they face significant limitations in domain-specific or proprietary contexts, where they are more likely to generate ``hallucinations'' responses that sound plausible but are inaccurate or fabricated \cite{ji2023survey}.



Lately, the cognitive and learning impacts of LLMs, search engines, or no tools (brain-only) have been studied with electroencephalography\footnote{
Electroencephalography (EEG) is a neurophysiological diagnostic test that measures and records the electrical activity of the brain using electrodes placed on the scalp.}, showing that LLM users exhibited the weakest neural connectivity and cognitive engagement, while brain-only participants had the strongest~\cite{kosmyna2025your}.
%
These findings raise serious concerns for fields like software modeling, where both students and educators increasingly rely on GenAI for tasks such as translating requirements, generating models, or designing curricula. While GenAI may offer convenience and initial support, its long-term impact on cognitive development and learning ownership seems troubling. Moreover, in many academic settings, the use of GenAI in exams and assignments is still considered a form of plagiarism \cite{hutson2024rethinking}, yet clear boundaries around acceptable use remain undefined. Without proper ethical guidelines, both learning integrity and teaching responsibility are at risk.

\subsection{Ethical Regulations}
In \cite{troquard2024social}, Troquard et al. discuss the \textit{SLEEC} rules, which stands for social, legal, ethical, empathetic, and cultural and emphasize the importance of identifying and integrating these dimensions in the context of human-AI interaction. As highlighted by emerging regulations such as the European AI Act, growing concerns around the societal and human impact of GenAI tools are moving into the public sphere. Beyond traditional software quality, values such as human well-being, social equity, and environmental sustainability are increasingly seen as critical for long-term resilience. In modeling educational settings, particularly for students, the integration of GenAI into assignments, lesson planning, modeling feedback, or even empirical research demands caution. The absence of regulation means outputs often diverge from expectations \cite{zawacki2019systematic}.

\subsection{Ethical Dimensions of GenAI in Software Modeling Education}
The three papers identified in our review do not address how students or educators should be guided through the ethical dimensions of GenAI in software modeling. Most modeling tools now integrate AI agents to generate diagrams from prompts. However, AI prompts, designed to facilitate human-AI conversation, often fall short in educational contexts. They are prone to plagiarism, ethical misuse, and limitations in handling abbreviations or unfamiliar terms. 

For example, is has been demonstrated that first names in prompts acting as demographic cues can significantly influence ethical decision-making by LLMs~\cite{berlincioni2024prompt}. This is an important consideration in modeling where prompt wording affects model structure. 

While ethics is crucial, incorporating it into already technical modeling courses is difficult~\cite{garrett2020more}.

\section{Related Work}
\label{sec:rw}
The integration of Generative AI tools such as ChatGPT into educational settings has sparked a growing body of research exploring their impact, benefits, and ethical considerations. Although these tools offer potential enhancements in learning and teaching processes, they also raise significant ethical concerns, especially in the context of software modeling education. 

\subsection{Use of GenAI in Software Modeling}
\label{subsec:GenAISoftwareModeling}
Camara et al.\cite{camara2023assessment} analyzed ChatGPT's ability to assist with UML modeling tasks, identifying both semantic and syntactic deficiencies, as well as risks regarding the consistency of generated responses. The authors suggest that while ChatGPT is not yet reliable for complex modeling tasks, it holds potential as an assistant tool in Model-Based-Software Engineering. They also acknowledge that the ethical implications of these tools are important and should be analyzed, although the paper focuses more on other technical aspects. Similarly, De Bari et al.\cite{DeBari24} evaluated whether LLMs can produce UML class diagrams comparable in quality to those created by humans. The study concludes that although the syntactic and pragmatic quality is acceptable, LLMs still have limitations in understanding domain concepts, affecting the semantic quality of the diagrams. However, the authors emphasize that the diagrams generated by these tools provide initial drafts that students and teachers can then refine. 

\subsection{Applied Ethics in Software Engineering Education}
\label{subsec:EthicsSE}
The integration of ethics into the training of software engineers has been the subject of study in recent years beyond the context of modeling. Pant et al.\cite{teachingsoftwareethicsfuture2023} proposed a teaching approach based on interactive ethical scenarios with the aim of promoting ethical decision making among students. Their study showed that this type of intervention significantly improves understanding of ethical dilemmas in real-word software development contexts.  The authors ultimately recommend the implementation of such programs to prepare future professionals to face ethical challenges in their professional practice.

Leça et al.\cite{leça2024responsibleaisoftwareindustry}, meanwhile, examined how software professionals deal with the principles of responsible artificial intelligence in their daily practice. They found that while aspects such as equity and inclusion are addressed, other key ethical principles, such as transparency and accountability, tend to be ignored or misinterpreted. This suggests an urgent need for deeper and more cross-cutting education in applied ethics.

\subsection{Ethical Dimensions of ChatGPT in Education}
\label{subsec:EthicalChatGPT}
Silva et al.\cite{silva2024chatgpt} conducted an experiment in which students used ChatGPT to program in C, revealing that while the tool is perceived as useful, it also raises related ethical issues. Specifically, the authors point out the following ethical problems: (a) plagiarism and code authorship, complicating the assessment of actual skills; (b) transparency and traceability, making it difficult to discern when a student has used AI and to what extent; (c) technological dependence, leading students to avoid independent reasoning; (d) inequality of opportunity, creating an imbalance between those with more technological resources and those without. Some of these ethical risks are shared by other authors in their work. For example, Weidinger et al.\cite{Weidinger2021} suggested six types of ethical risks, including (1) discrimination, exclusion and toxicity, (2) information risks, (3) harms from misinformation, (4) malicious uses, (5) harms from human-computer interaction and (6) harms from automation, access and the environment.

\subsection{Practical and Ethical Challenges}
\label{subsec:EthicalChallenges}
The use of ChatGPT and other generative models poses not only ethical risks but also multiple practical challenges for their effective integration into educational contexts. Li et al. \cite{li2024mitigating} proposed strategies to mitigate the dependence of students on ChatGPT, such as encouraging critical thinking so that they can evaluate AI-generated content from an ethical and values-based perspective and limiting its use in assessments. The authors highlighted in their work that AI improves the sophistication of the educational environment, while also posing numerous ethical and moral challenges. Among these ethical challenges, they point out (a) the erosion of personal autonomy, (b) emotional distancing, (c) alienation in student development, and (d) risks to personal privacy. 

According to Yang et al.\cite{Yan_2023}, a key barrier to the ethical use of LLMs in education is the lack of transparency in terms of the origin of the generated content. In conceptual modeling scenarios, this opacity affects the traceability of decisions in model design, compromising academic accountability, and hindering the assessment of individual competencies.

\subsection{Regulatory Framework and Ethical Governance}
\label{subsec:framework}
In light of the unstoppable advance of GenAI, it is necessary to establish clear regulatory frameworks to govern its use in educational contexts. In this regard, Wu et al.\cite{articleWu2024} explore how major US universities are addressing the ethical implications of GenAI through decentralized, yet coordinated governance structures. Ethical concerns are embedded in the guidance provided to different stakeholders, highlighting transparency, accountability, and responsible use. Rather than imposing rigid rules, the guidelines promote ethical awareness, with both teachers and students participating in the design of usage policies. This participatory approach allows tools to be adapted to the specific realities of the classroom, minimizing ethical risks and maximizing pedagogical benefits. However, it is important to highlight the importance of linking institutional guidelines with international standards such as those proposed by UNESCO\cite{allahRakha2024unesco} or the OECD\cite{oecd2019ai}, to avoid localist or inconsistent approaches. Without regulatory convergence, academic environments will continue to produce fragmented models of GenAI adoption without solid ethical criteria. 

Taken together, these studies show that while interest in educational applications of GenAI is growing, there is still a limited and fragmented discussion of its ethical implications, particularly in the specific domain of software modeling. Our work aims to address this gap by focusing on the need for structured ethical reflection within AI-assisted modeling education.

\section{Conclusions}
\label{sec:conclusions}
While ethical issues are starting to be discussed and acknowledged in broader software engineering educational settings and programming contexts, modeling education remains largely untouched. 
We argue that this discussion must go beyond the broader context of CS/SE education and specifically address modeling education. Unlike programming assignments, where generative AI has already been widely studied, modeling tasks require a different set of skills like abstraction, system thinking, and design reasoning that are uniquely challenged by GenAI. Moreover, generative AI tends to perform worse in domain-specific modeling tasks compared to mainstream coding tasks, raising distinct risks for students who may develop misleading intuitions about model correctness, completeness, or fairness. Hence, ethical concerns in modeling education deserve a dedicated perspective.

Our findings expose a significant research gap: a lack of ethical maturity within modeling education. As AI technologies, particularly Generative AI, become increasingly embedded in modeling tools and pedagogical practices, the absence of structured ethical guidance becomes more concerning. From issues of fairness, bias, and inclusion to questions around cognitive impact, authorship, and sustainability, these dimensions are either minimally addressed or completely overlooked.

We advocate for a more intentional and structured integration of ethical reflection into AI-supported modeling education. This means not only raising awareness among students and educators but also embedding ethical considerations directly into curricula, tools, and assessment frameworks. Institutions and researchers must go beyond compliance-level acknowledgments and engage critically with the implications of using GenAI in formative learning environments. Without this, we risk fostering a generation of modelers who are technically proficient but ethically unprepared to navigate the complex realities of AI-assisted software engineering.
Looking forward, we see several important open research questions that could guide future work:
\begin{itemize}
    \item \textbf{Explainability:} How can GenAI-generated models be made explainable to students so that they understand the reasoning behind the AI’s outputs?

    \item \textbf{Bias detection:} What methods can be developed to identify and mitigate hidden biases in AI-generated models, particularly those introduced through prompts or training data?

    \item \textbf{Pedagogical integration:} How can ethical reflection be systematically integrated into already technical modeling courses without overburdening the curriculum?

    \item \textbf{AI reliance:} To what extent does the use of GenAI in modeling education risk reducing students’ independent problem-solving abilities, creativity, and critical thinking?

    \item \textbf{Evaluation and scaffolding:} What teaching strategies, scaffolding tools, or assessment methods can support students in responsibly using GenAI for modeling tasks?

\end{itemize}
Answering these questions will be essential to ensure that the integration of GenAI into modeling education not only enhances learning but also prepares students to critically and responsibly engage with these technologies in their future professional practice.

\bibliographystyle{IEEEtran}
\bibliography{IEEEabrv,literature}

\end{document}